\documentclass[11pt]{article}
\usepackage[margin=0.75in]{geometry}
\usepackage{titling}
\usepackage{setspace}
\usepackage[numbers]{natbib}
\usepackage{fancyhdr}
\usepackage[hidelinks]{hyperref}
\usepackage{authblk}
\usepackage{IEEEtrantools}

\renewcommand\Affilfont{\itshape\small}
\makeatletter
\renewcommand\AB@affilsepx{\quad\protect\Affilfont} % put affiliations into one line
\makeatother

\makeatletter
\renewcommand\maketitle{
{\raggedright % Note the extra {
\begin{center}
  {\setstretch{1.2}\Large \bfseries \@title\par}
  \vspace{2ex}
{ \@author}\\[2ex]
\@date\\[3ex]
\end{center}}} % Note the extra }
\makeatother

\title{Recommendations to clarify NASA open source requirements}
\author[1]{John D. Haiducek}
\affil[1]{U.S. Naval Research Laboratory}
\author[2]{Thom R. Edwards}
\affil[2]{Technical University of Denmark (former)}
\author[1]{Wade Duvall}
\author[3]{Sarah R. Cannon}
\affil[3]{Independent researcher}
\author[4]{Kai Germaschewski}
\affil[4]{University~of~New~Hampshire}
\author[1]{Jason E. Kooi}
\date{29 September 2021}
\begin{document}

\maketitle

\bstctlcite{IEEEexample:BSTcontrol}

\begin{abstract}
  The software community has specific definitions for terms such as ``open source software,'' ``free software,'' and ``permissive license,'' but scientists proposing software development efforts to NASA are not always knowledgeable about these definitions. Misunderstandings about the meaning of these terms can result in problems of fairness with solicitations, because scientists who interpret the terms differently than NASA intends may either needlessly limit the scope of their proposed work, or unwittingly propose work that does not comply with software licensing requirements. It is therefore recommended that NASA adopt definitions of the above terms that are in line with software community usage, that these definitions be communicated as part of solicitations to ensure a common understanding, and that proposals be required to identify what software licenses the proposers expect to use.
\end{abstract}

\subsection*{Description of the problem}

In recent years NASA has established policies designed to encourage development of and participation in open source software (OSS). However, confusion over OSS requirements has sometimes led to problems. Stakeholders including grant applicants and reviewers, legal teams, software developers, and software users all have different understandings of what they think qualifies as OSS, and these differences can lead to confusion about what is required to satisfy OSS requirements. Improving clarity within NASA and among NASA stakeholders as to the meaning of OSS and related terms would improve NASA's ability to assess compliance with requirements related to OSS. This would also increase fairness in the solicitation process, by establishing clear expectations around software licensing.

In the software community the term ``open source'' refers to software whose source code is distributed publicly under a license that satisfies a set of criteria designed to ensure that the software can be freely used, modified, and shared. These criteria are laid out in the open source definition provided by the Open Source Initiative (OSI) \citep{OpenSourceDefinition}. OSS that satisfies the slightly more stringent definition of ``free software'' from the Free Software Foundation (FSF) \citep{WhatFreeSoftware} is sometimes called ``free/libre and open source software'' (FLOSS or FOSS) \citep{WhyOpenSource,FLOSSFOSSGNU}. Both definitions are quite detailed in order to address various issues that have come up with software licenses over the years. Some NASA policies also mention a subset of OSS licenses called ``permissive'' licenses \citep{ESDSOpenSource}. A ``permissive'' license is an open source license that allows the creation of proprietary derivative works (as opposed to a ``copyleft'' license that requires derivative works use the same license as the original) \citep{WhatCopyleftGNU,FrequentlyAnsweredQuestions}.

Some NASA documents and policies have acknowledged the OSI and FSF definitions as widely accepted \citep{ESDSOpenSource,Moran2003DevelopingOpenSource}, but NASA does not always use and apply these definitions consistently. Moreover, many scientists mistakenly understand the term ``open source'' to mean simply that source code is available to the public. As a result, some software products developed by scientists are advertised as ``open source'' even though their licenses violate one or more of the ten criteria of the OSI definition. Common examples include licenses that prohibit sale of the software (violating criterion \#1, ``Free redistribution,'' which requires in part that the license allow sale of the software as part of an aggregate product) \citep{OpenSourceDefinition}, or that prohibit commercial use (violating criterion \#6, ``No discrimination against fields of endeavor'').

Mislabeling software as open source when it does not meet the OSI or FSF definitions can encourage violations of non-OSS licenses. Users and developers who are told that a software product is OSS may choose to use or develop the software on the assumption that the software license complies with the OSI and/or FSF definition. The developer may have greater difficulty enforcing the license terms as a result. Users may invest time and effort into the software based on the understanding that it is OSS, only to read the license later and find it unsuitable to their purposes. Inconsistent or confusing licensing can mislead users and harm the reputation of both the developer and NASA. Having a common understanding of what is meant by terms like ``open source'' would reduce these kinds of problems.

Although source code availability can be beneficial by itself, software that does not fully meet both the OSI and FSF definitions misses out on many of the advantages of OSS. License restrictions that satisfy only the OSI definition have been identified as a barrier to adoption for NASA-developed software \citep{NationalAcademiesofSciencesEngineeringandMedicine2018OpenSourceSoftware,Beyer2018NoNOSAYes,VariousLicensesComments}. These restrictions have resulted in some software products being rejected for packaging in Linux distributions \citep{ReviewRequestCdf,LicensingMainFedora}. A simple way to avoid these kinds of problems is to use one of the licenses already approved by both OSI and FSF \citep{VariousLicensesComments,OpenSourceLicenses}, rather than trying to write a custom license. Some U.S. government programs have taken the approach of requiring OSI-approved licenses for developed software \cite{NationalAcademiesofSciencesEngineeringandMedicine2018OpenSourceSoftware}.

In NASA solicitations that include an OSS requirement, misunderstandings around terms such as ``open source'' and ``permissive'' can cause fairness issues. Proposal teams that interpret the OSS requirements more strictly than NASA does may limit the scope of their proposed work as a result, putting themselves at a disadvantage relative to other teams. On the other hand, proposal teams that interpret the OSS requirements more loosely than NASA does may find their proposals rejected. If NASA were to adopt the OSI and FSF definitions and state in solicitations that these definitions apply, this would make clearer to proposal teams what is required to comply with policies.

A related problem is that proposal teams may commit generally to open source development, but without specifying what license they plan to use. If the proposal team uses a looser definition of ``open source'' than NASA does, they might write a proposal that appears to be in compliance, when the team in fact plans to use a software license that is not consistent with the standards expected by NASA. If the license applies to third party software, the proposal team might not have the authority to change it. This situation could create problems for both NASA and the proposal teams, which might not become apparent until after the work has begun.

\textbf{Establishing common ground as to the meaning of terms related to OSS, and increasing clarity of communications around software licensing, would benefit NASA and NASA-funded scientists.}

\subsection*{Recommendations}

To ensure a common understanding around terms related to OSS, and to simplify assessment of OSS-related requirements, it is recommended that NASA do the following:

\begin{enumerate}
\item \textbf{Explicitly adopt definitions for ``open source'' and ``permissive'' that are in line with the definitions provided by OSI and FSF.}

  This clarifies NASA's intent when using these terms. Aligning NASA's usage with that of the software community reduces the chances of misunderstandings.

\item \textbf{Include definitions of ``open source,'' and ``permissive'' in solicitations where these terms are used.}

  This ensures that proposers are aware of NASA's intended meaning when these terms are used, and reduces the chances that a proposer will unwittingly rely on a different definition.

\item \textbf{Provide clear guidance when NASA has licensing requirements that differ from the standard definition of ``open source.''}

  In some solicitations NASA may choose to require that source code be freely distributed, but not require full compliance with the OSI or FSF definitions. For instance, NASA might allow ``non-commercial'' licenses or ones that restrict sales. In such cases, the solicitation should specify what kinds of license terms will be allowed or disallowed.
  \filbreak

\item \textbf{Require proposals include a list of software products the team expects to create or modify, and the expected license for each.}

  This enables NASA to better assess whether software licensing requirements have been met.

\item \textbf{Encourage the use of OSI or FSF approved licenses rather than custom licenses.}

  This reduces the effort required to assess compliance, by reducing the need to read each license, and lowers barriers to adoption of software outside NASA. It may also reduce the effort required by proposers, by reminding them that they can choose an existing license rather than writing their own.

\end{enumerate}

\subsection*{Benefits to NASA and the science community}

Making clearer and more consistent communications about open source licensing would place NASA in a leadership position within the U.S. government in terms of open source software policy, since other agencies currently also exhibit inconsistencies in this area \citep{NationalAcademiesofSciencesEngineeringandMedicine2018OpenSourceSoftware}. Better communications about licensing in solicitations will increase fairness in the selection process, and clarify expectations once projects are underway. Aligning NASA's usage of software licensing terms with those of the software community will help reduce confusion at the solicitation phase between proposers, reviewers, and NASA. Although NASA might not require open source development in every case, this ensures that what NASA claims as open source will be accepted as such outside NASA. Encouraging the use of OSI and FSF approved licenses will reduce barriers to adoption of NASA-funded software outside the agency by making it easier for outside users and developers to understand and accept the license terms. This will enable greater involvement of outside developers and increase the availability of the software to users by enabling third party distribution. Both NASA and the nation will benefit as a result.

\subsection*{Acknowledgements}

Thanks to Meghan Burleigh for her assistance in the editing process.

{\footnotesize
  \bibliography{IEEEtranBSTCTL,sources}}

\end{document}